\documentclass[pra
,preprintnumbers,amsmath,amssymb]{revtex4}

\usepackage{graphicx}
\usepackage{dcolumn}
\usepackage{bm}

\begin{document}

\title{Tunneling of polarized fermions in 3D double wells}

\author{T. Macr\`{i}}
\affiliation{Max Planck Institute for the Physics of Complex Systems, 
N\"othnitzer Str. 38, 01187 Dresden, Germany}
\affiliation{SISSA and INFN, Sezione di Trieste, via Bonomea 265, I-34136, 
Trieste, Italy}

\author{A. Trombettoni\footnote{email address: andreatr@sissa.it}}
\affiliation{CNR-IOM DEMOCRITOS Simulation Center, Via Bonomea 265, I-34136 Trieste, Italy}
\affiliation{SISSA and INFN, Sezione di Trieste, via Bonomea 265, I-34136, 
Trieste, Italy}


\begin{abstract}
We study the tunneling of a spin polarized Fermi gas in 
a three-dimensional double well potential, focusing on the time dynamics 
starting from an initial state in which there is an imbalance 
in the number of particles in the two wells. 
Although fermions in different 
doublets of the double well tunnel with different frequencies, we point out 
that (incoherent) oscillations of a large number of particles can arise, 
as a consequence of the presence of transverse degrees of freedom. Estimates 
of the doublet structure and of the occupation of transverse eigenstates 
for a realistic experimental setup are provided.   
\end{abstract}

\maketitle

\section{Introduction}

Tunneling of particles through wells and barriers is a distinctive  
property of quantum mechanics, commonly exploited in the realization 
of solid-state devices \cite{BARONE82,ROY86}. The experimental realization 
and manipulation of ultracold atoms \cite{PITAEVSKII03,PETHICK08} 
in double- and many- well potentials makes possible to study tunneling dynamics of 
fermions and bosons in an highly controllable setup, in which 
it is possible to tune the geometrical properties of the 
wells, and then the tunneling rate of the particles. 
For ultracold bosons, the coherent Josephson oscillations of 
a Bose-Einstein condensate in a double well potential were discussed
\cite{SMERZI97,ZAPATA98} and subsequently experimentally observed 
\cite{ALBIEZ05,LEVY07,LEBLANC11}; the dynamics of bosonic squeezed states 
has been as well investigated \cite{ESTEVE08}. 
For large enough barriers between the two wells, 
a two-mode ansatz can be used to describe the tunneling dynamics: for 
bosonic condensates, 
the two-mode equations can be mapped in a non-rigid pendulum \cite{SMERZI97} 
and can also be generalized to many-well potentials 
\cite{TROMBETTONI01,MORSCH06}. The dynamical properties of cold 
bosons in double- and many- wells potential, as well as 
double-well arrays, in presence of driving time-dependent 
modulations has been also subject of intense study 
\cite{ECKARDT05,GRAEFE06,LIGNIER07,KIERIG08,ECKARDT09,ZENESINI10,MORALES10,NESTERENKO10,STRUCK11,ESMANN12}.

On the other hand, the study of ultracold Fermi gases has known a great advance through 
the last years \cite{GIORGINI08,BLOCH08,KETTERLE08} and the impressive experimental progresses in their control 
make Fermi gases very promising to realize ultracold fermionic junctions and 
to study tunneling phenomena. Due to the possibility to 
use optical lattices \cite{MODUGNO03,CHIN06,STOFERLE06,SCHNEIDER08} 
and to tune the interspecies interaction through Feshbach resonances \cite{PITAEVSKII03,PETHICK08}, the fermionic tunneling 
can be studied in situations having a direct counterpart in solid-state 
tunnel junctions. When the Fermi gases in different wells are superfluid 
(due to attractive interaction among species) one has the ultracold fermionic equivalent of a superconductor-insulator-superconductor junction \cite{BARONE82}, having coherent tunneling: 
theoretical studies of tunneling of fermionic 
superfluids through barriers \cite{SPUNTARELLI07,WATANABE09,SPUNTARELLI10} and in double- and many- well potentials 
\cite{WOUTERS04,TEMPERE05,SALASNICH08,WATANABE08,XUE08,ANCILOTTO09,ADHIKARI09} 
have been reported in literature, as well as the study of the 
internal Josephson tunneling between different species \cite{PARAOANU02} 
(see more references in \cite{SPUNTARELLI10}). 

At variance, one can also realize the ultracold atomic counterpart 
of one or more {\em normal} tunnel junctions \cite{ABRIKOSOV88} 
when the temperature of the Fermi mixture in a double- or many- well potential 
is above the superfluid critical temperature 
(in presence of attractive interactions) or, even at zero temperature, 
when the Fermi gases in the wells are in the normal state. 
The latter situation can be 
obtained polarizing above a critical threshold a two-component Fermi mixture 
\cite{ZWIERLEIN06,PARTRIDGE06} or simply using spin polarized fermions 
\cite{PEZZE04,SALASNICH10}. 
In \cite{PEZZE04} the center-of-mass motion 
of a polarized Fermi gas in a combined periodic and harmonic potential was 
theoretically and experimentally investigated, showing an insulating regime 
when the Fermi energy lies into the bandgap of the lattice: working 
in tight-binding approximation, the dynamics of the polarized gas 
was studied solving the classical Liouville equation, allowing for the 
characterization of the different regimes of the center-of-mass dynamics 
of the 3D Fermi gas \cite{PEZZE04,PEZZE04-2}. 
Rabi oscillations of a degenerate fermionic gas 
in a double well potential has been studied in \cite{CHWEDENCZUK09} 
in relation to the possibility to exploit them for the 
interferometric measurement of external forces at micrometer length scales. 
The tunneling dynamics of interacting bosons in a 1D double well,  
from weak interactions to the fermionization (Tonks-Girardeau) limit,  
was studied in \cite{ZOLLNER08}. 

In this paper we study the tunneling 
dynamics of an ideal (spin polarized) fermionic gas 
in double well potentials. Of course, unlike condensed Bose gases where 
a large number of particles are in the same state moving coherently, 
for polarized fermions the dynamics displays in general an incoherent 
motion of the particles in the external potential. 
The situation is analogous to the incoherent tunneling of 
electrons between normal metals in the approximation where they 
are considered free particles: 
the application of a constant voltage creates a difference in 
the chemical potential and the more energetic 
electrons can tunnel through the barrier. For Fermi gases an imbalance 
in the chemical potential can be either obtained by 
making asymmetric the double well potential (with the two wells 
having different potential minima) 
or creating an imbalance in the number of particles between the two wells 
of the system at the initial time: in this paper we will focus 
on the latter situation. 
Because of Pauli principle, polarized fermions do not interact 
in s-wave and at $T=0$ they occupy all the doublets of the 
double well potential up to the Fermi energy 
(the splitting of doublets increases when 
the energy of the doublets increases).
Since in general fermions in different doublets 
tunnel with different frequencies, then a dephasing in the current flowing 
among the wells may arise. This is what happens in the 1D case 
\cite{SALASNICH10}: the transverse degrees of freedom are frozen, since 
the distances between different doublets of the 1D double well potential 
are much smaller than the confining frequencies in the 
transverse directions. The tunneling dynamics then shows
strongly aperiodic spatio-temporal patterns \cite{SALASNICH10}: for large 
number of fermions, current oscillations are practically washed out. 

However, for a relatively large number of fermions (say $N\gtrsim 10^3$), 
the validity of 1D limit requires very large confining transverse 
frequencies: e.g., $\omega_\perp/2\pi \, \gtrsim 1000kHz$ 
for $^{40}K$ atoms with typical experimental values
for the potential.
Indeed, as we will discuss in Section 
II, if one has an energy barrier $V_0$ between the wells and the minima of the 
wells are at distance $\lambda/2$, then  
the energy difference between the doublets is 
$\sim \hbar k \sqrt{V_0/m}$ (where 
$k=2 \pi/\lambda$ and $m$ the mass of the fermionic atom). The condition 
for which the transverse degrees of freedom are frozen is then 
$N\hbar k \sqrt{V_0/m} \lesssim \hbar \omega_\perp$. 
For a realistic double well potential 
\cite{ALBIEZ05} one has $V_0/h \sim 1 kHz$ and  
$\lambda \sim 2 \mu m$: with $m$ the mass of potassium ($^{40}K$) atoms, 
one gets $\omega_\perp / (2\pi N) \gtrsim 1kHz$, corresponding to transverse 
frequencies $\omega_\perp/2\pi\, \gtrsim 1000kHz$. 
One then sees that for typical transverse 
frequencies one has to take into account the transverse degrees of freedom. 

In the following we consider a realistic 3D double well, obtained 
superimposing an harmonic confinement with a periodic potential: 
properly tuning the parameters, one can have two wells much more 
populated than the others \cite{ALBIEZ05}. We provide estimates 
of the tunneling rates and of the occupation of the transverse degrees 
of freedom. In particular, we show that it is realistically possible 
to have only the first doublet occupied,
showing the role played by the transverse degrees of freedom 
in providing a reservoir to store particles and resulting 
in oscillations of a large number of particles with the same frequency. 
We stress that these are incoherent single-particle oscillations. Then, 
if the 1D condition is violated the simple 
observation of a sinusoidal current cannot discriminate between 
incoherent and coherent tunneling dynamics (the latter being expected 
when the weakly coupled Fermi gases are superfluids).

The plan of the paper is as follows: 
in Section II we analyze the structure of levels for a
3D Fermi gas in a double well potential obtained by 
the superposition of an harmonic confinement 
and a 1D optical lattice. In this Section we also provide 
estimates for the number of doublets 
and the occupation of transverse eigenstates in a realistic setup. 
In Section III we discuss the equations of motion for the dynamics in a 
3D double well.
In Section IV we study the dynamics of the Fermi gas in the
3D double well potential described in Section II, showing 
how oscillations characterized by a single frequency can arise 
and studying deviations due to the occupation of higher doublets. 
Finally in Section V we present our conclusions, while 
in Appendix A we report semiclassical estimates 
of the energy and the splitting of the doublets.

\section{Polarized fermions in a 3D  double well potential}

In this Section we study the structure of levels and their filling by a polarized  Fermi 
gas in a 3D double well potential. We focus on the double 
well potential obtained by superimposing an harmonic confinement and 
a 1D periodic potential: when the energy of two 
minima of the periodic potential is significantly lower than the energies 
of other minima, due to the presence of the harmonic trap, then 
these wells are much more populated than the others and one practically 
has a double well. This way to realize the double well 
has been used to study the tunneling 
dynamics of ultracold bosons 
in a double well potential in \cite{ALBIEZ05,ESTEVE08} and could be used also 
for Fermi gases. The energy barrier and the distance 
between the wells are controlled by acting on the parameters of the optical 
lattice \cite{MORSCH06}, while the number of doublets under barrier 
depends also on the trapping frequencies of the harmonic confinement. 
In the following we will discuss how the fermions fill the doublet structure 
for a realistic 3D double well, 
keeping into account the transverse eigenstates. 

\begin{figure}[t]
\centering
\includegraphics[width=6.0cm,angle=0,clip]{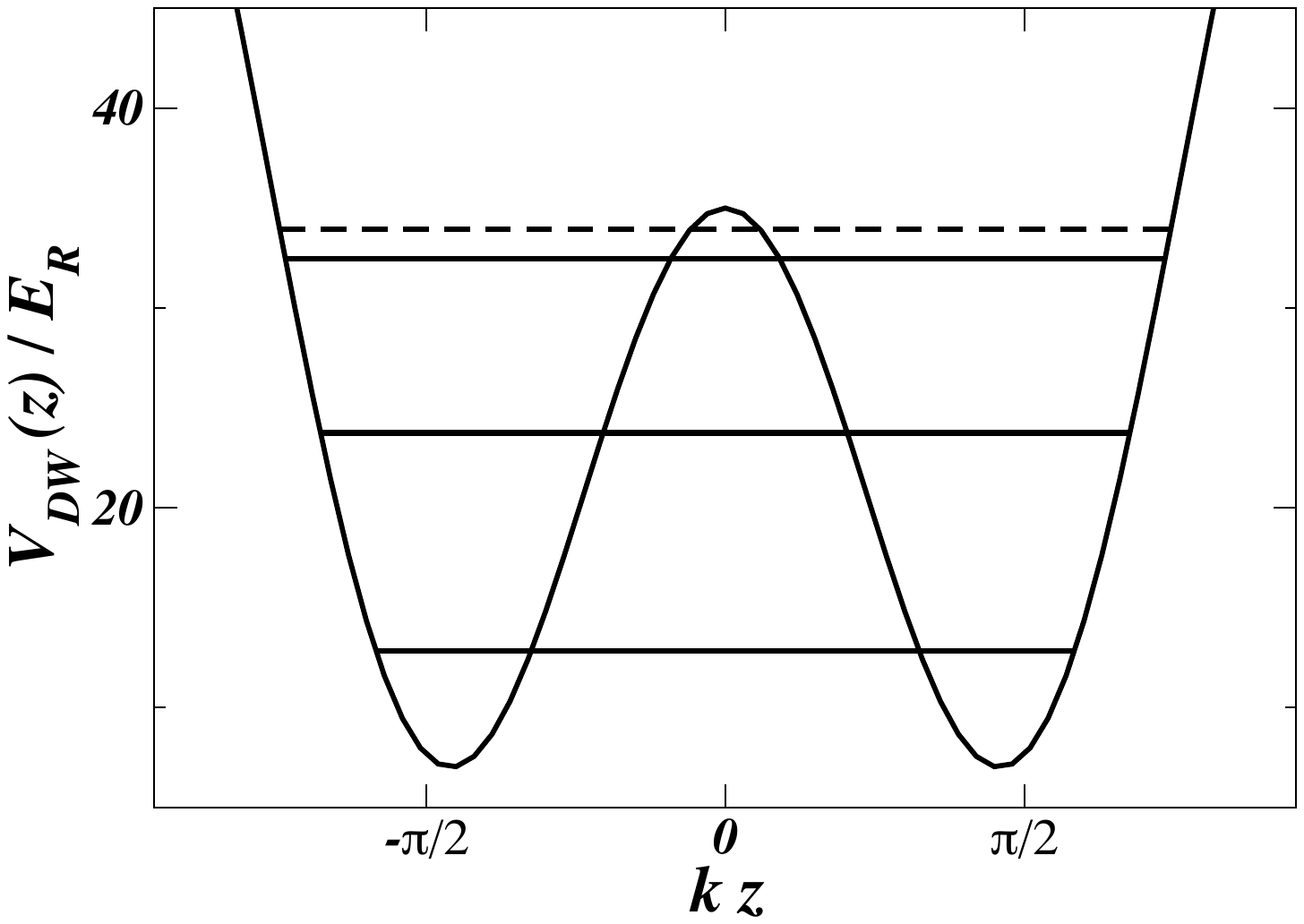}
\hspace{5mm}
\includegraphics[width=6.0cm, angle=0]{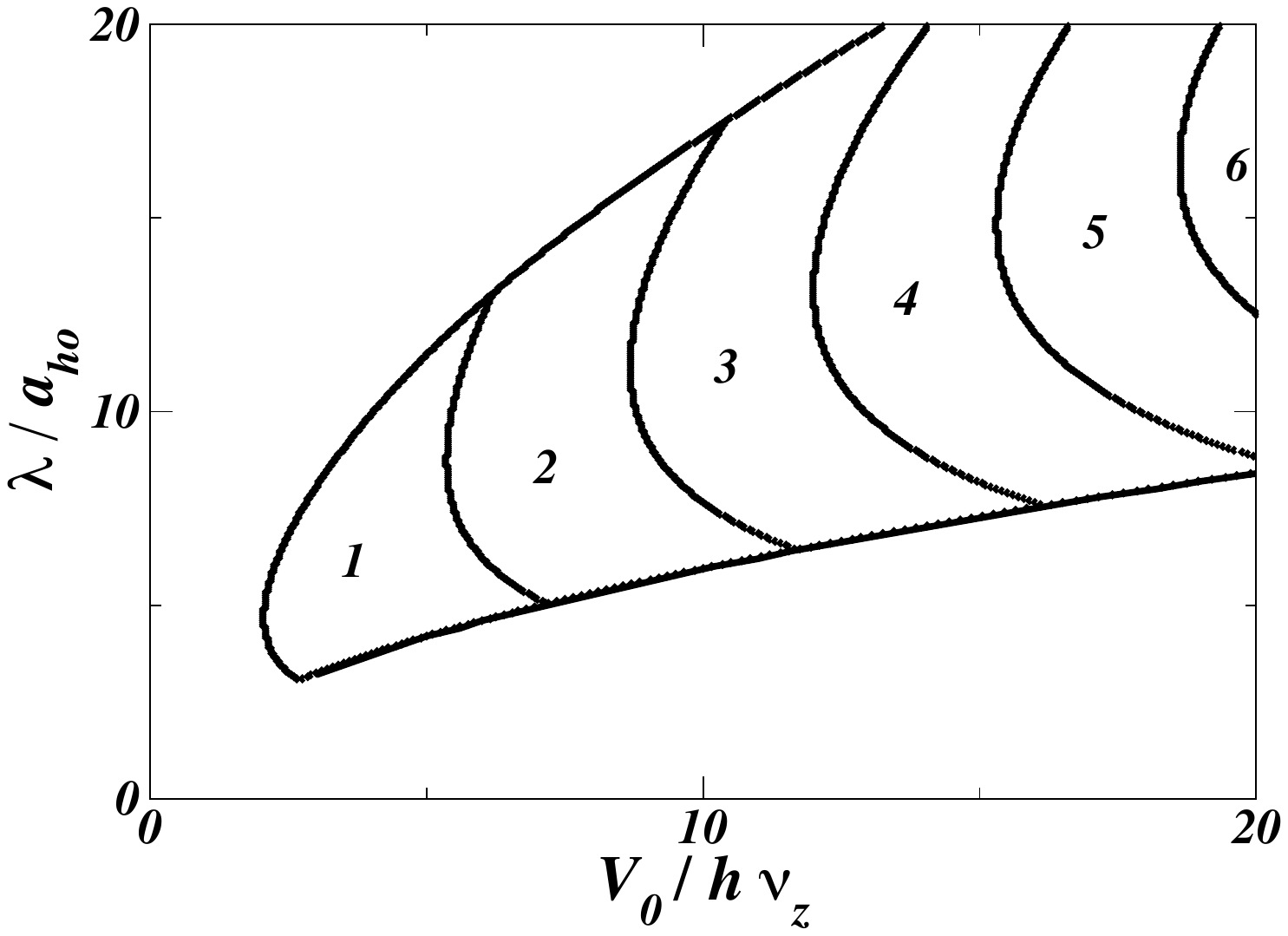}
\caption{(Left) Plot of the potential $V_{DW}(z)$. Here we use  the values $\omega_z=2\pi \cdot 78 $ Hz, 
$\lambda=  15 \mu m$ and $V_0/h=35 \text{E}_R/h= 4.8 \text{kHz}$ corresponding to $s=35$ and $A=0.12$. The double well has
three doublets under the central barrier; the upper solid line correspond to the symmetric energetic level of the third doublet; 
the dashed one to the antisymmetric level. For the first two doublets 
(solid lines) it is not possible to distinguish among
symmetric and antisymmetric levels because are too close in energy.
(Right) Number of doublets as a function of the strength $V_0$ and the wavelength of the
periodic potential expressed, respectively, in terms of $h \nu_z$ and $a_{ho}$. }
\label{doublets}
\end{figure}

The trapping potential reads
\begin{equation}
V(\mathbf{x}) = \frac{1}{2} m \omega_\perp^2 (x^2 + y^2) + V_{\text{DW}}(z),
\label{V_TOT}
\end{equation}
where the double well potential along the $z$-axis, 
$V_{\text{DW}}(z)$, has the form 
\begin{equation} \label{double_potential}
V_{DW}(z) = \frac{1}{2} m \omega_z^2 z^2 + V_0 \cos^2(k z)
\end{equation}
[see Fig.\ref{doublets} (left)]. 
In (\ref{double_potential}), the 1D periodic potential is 
created by an optical lattice made of two counterpropagating laser beams: 
it is $k=2 \pi/\lambda$ where $\lambda = \lambda_{laser} \sin{(\theta/2)}$, 
$\lambda_{laser}$ being the wavelength of the lasers and $\theta$ 
the angle between the counterpropagating laser beams \cite{MORSCH06}. 
The spacing in the lattice is $\lambda/2$, and $V_0$ is proportional 
to the power of the laser; moreover, $\omega_z$ and $\omega_\perp$ are respectively 
the axial and transverse frequencies of the harmonic confinement. 
Typical experimental numbers are $\lambda \sim 1-10 \mu m$ and 
$\omega_z/2\pi \sim 10-100Hz$: for $\lambda \sim 10 \mu m$, 
the energy barrier (needed to have tunneling dynamics) 
is $V_0 / h \gtrsim 500 Hz$ 
\cite{ALBIEZ05}. In the following we set $V_0=s E_R$, where 
$E_R=\frac{\hbar^2 k^2}{2m}$ is the recoil energy, and 
we also introduce the ratio 
\begin{equation}
A=\frac{a_{ho}}{\lambda}
\label{A}
\end{equation} where
$a_{ho}=\sqrt{\hbar/m \omega_z}$ is the harmonic oscillator 
length in the $z$-direction. 

For suitable ranges of the parameters, 
the potential (\ref{V_TOT}) is to a very good approximation
a double well potential [see Fig.\ref{doublets} (left)]. 
For large enough barriers between the two wells, 
a two-mode ansatz can be used to describe the tunneling dynamics of ultracold 
bosonic condensates \cite{SMERZI97,RAGHAVAN99,ANANIKIAN06}: 
in the two-mode approximation, 
the dynamics involves the lowest doublet.  
Similarly, in a periodic potential the dynamics of ultracold bosons  
in the tight-binding approximation involves only the lowest band 
\cite{TROMBETTONI01,MORSCH06}. Quasi-equilibrium mixtures of itinerant and 
localized bosons in optical lattices were studied in \cite{YUKALOV11-2}. 
At variance, polarized fermions occupy all the doublets up to the Fermi energy.

The potential $V_{DW}(z)$, rescaled in terms of $E_R$,
reads:
\begin{equation} \label{V_DW}
\frac{V_{DW}(z)}{E_R} = s\cos^2{(kz)} +\frac{1}{(2 \pi A)^4} (kz)^2.
\end{equation}
One easily sees that for large barriers, $s \gg 1$, 
the first two minima of Eq.(\ref{V_DW}) \cite{NOTE1} are very close to
$kz=\pm \frac{\pi}{2}$. Expanding the 
potential around these minima we get the effective
trapping frequency 
\begin{equation} \label{omega_tilde}
\tilde\omega=\frac{2\sqrt{s} E_R}{\hbar}=k \sqrt{\frac{V_0}{m}}:
\end{equation}
the energy difference between (neighbour in energy) doublets is 
then $\sim \hbar \tilde{\omega}$. This frequency then allows 
to estimate the mean energy and the number of doublets under the potential
barrier $V_0$. In general the splitting of doublets increases when 
the energy of the doublets increases. 
A first estimation of the splitting of each doublet can
be performed by means of a semiclassical computation \cite{LANDAU81}: 
in Appendix A we report semiclassical estimates 
of the energy and the splitting 
of the doublets.

In Fig.\ref{doublets} (right)  
we plot the number of doublets for the potential of 
Eq.(\ref{V_DW}) for different values of $V_0$ (in units 
of $\hbar \omega_z$) and $\lambda=2\pi/k$ 
obtained from the solution of the time-independent Schr\"odinger equation. 
We observe that for a fixed ratio $a_{ho}/\lambda$ 
there is a finite range of values of $V_0$: for $V_0$ smaller than 
a critical value $V_0^{(min)}$ the 
central barrier is not high enough to contain any doublet.
For large values of $V_0$, the nearest energy minima (located, for 
$s \gg 1$, close to $kz=\pm \frac{3\pi}{2}$) have an energy 
smaller than the potential energy at the top of the barrier, which 
is $sE_R$, and then the double well structure disappears. 
From the condition $V_{DW}(z=\pm \, 3\pi/2k)>V_{DW}(0)$ 
one gets the maximum value $V_0^{(max)}$, which is given by 
$V_0^{(max)} \approx 9E_R / 64 \pi^2 A^4=9 m \omega_z^2 \lambda^2/32$.

We now turn our attention to the evaluation of 
how the fermions fill the energy levels, in order to be able to make contact with
with typical ultracold atom experiments, which involve $\sim \, 10^3 \, - 
\, 10^5$ particles.
Since typically $\tilde{\omega}\gg \omega_z$, then the system is 
essentially 1D when $N\tilde{\omega} \ll \omega_\perp$,
where $N$ is the total number of particles. 
In this case fermions begin to occupy the levels of the
double well potential along $z$.
However for typical experimental values [see Fig.\ref{doublets} (right)], 
in a 1D setting the  
Fermi energy easily exceeds the barrier
even for a number of fermions of a few tenth.

The opposite 3D limit is obtained when $N \omega_\perp \ll \tilde{\omega}$.
In this limit, all the particles stay in the first doublet, 
occupying the transverse eigenstates associated to the 
doublet eigenstates. More precisely, 
to accommodate all the particles in the first doublet, 
two conditions must be satisfied at the same time: 
the energy difference between 
the first and the second 
doublet states ($\sim \hbar \tilde{\omega}$) 
must be much larger than the energy of the transverse modes 
($\propto \hbar \omega_\perp$) and 
the transverse trapping frequency $\omega_\perp$ must be much 
smaller than $\omega_z$. 
Indeed, if $\omega_\perp \sim \omega_z$, then 
the two conditions $\tilde{\omega} = k \sqrt{V_0/m} 
\gg \omega_\perp \sim \omega_z$ and 
$V_0<V_0^{(max)}=(9/32) m \omega_z^2 \lambda^2$ 
(needed to have a double well structure) cannot be at the same satisfied. 
Finally, we notice that if we relax the previous conditions then 
higher doublets can be occupied and situations intermediate between the 1D and 
3D limits can be explored.

\begin{figure}[ht]
\centering
\includegraphics[width=7.cm,angle=0,clip]{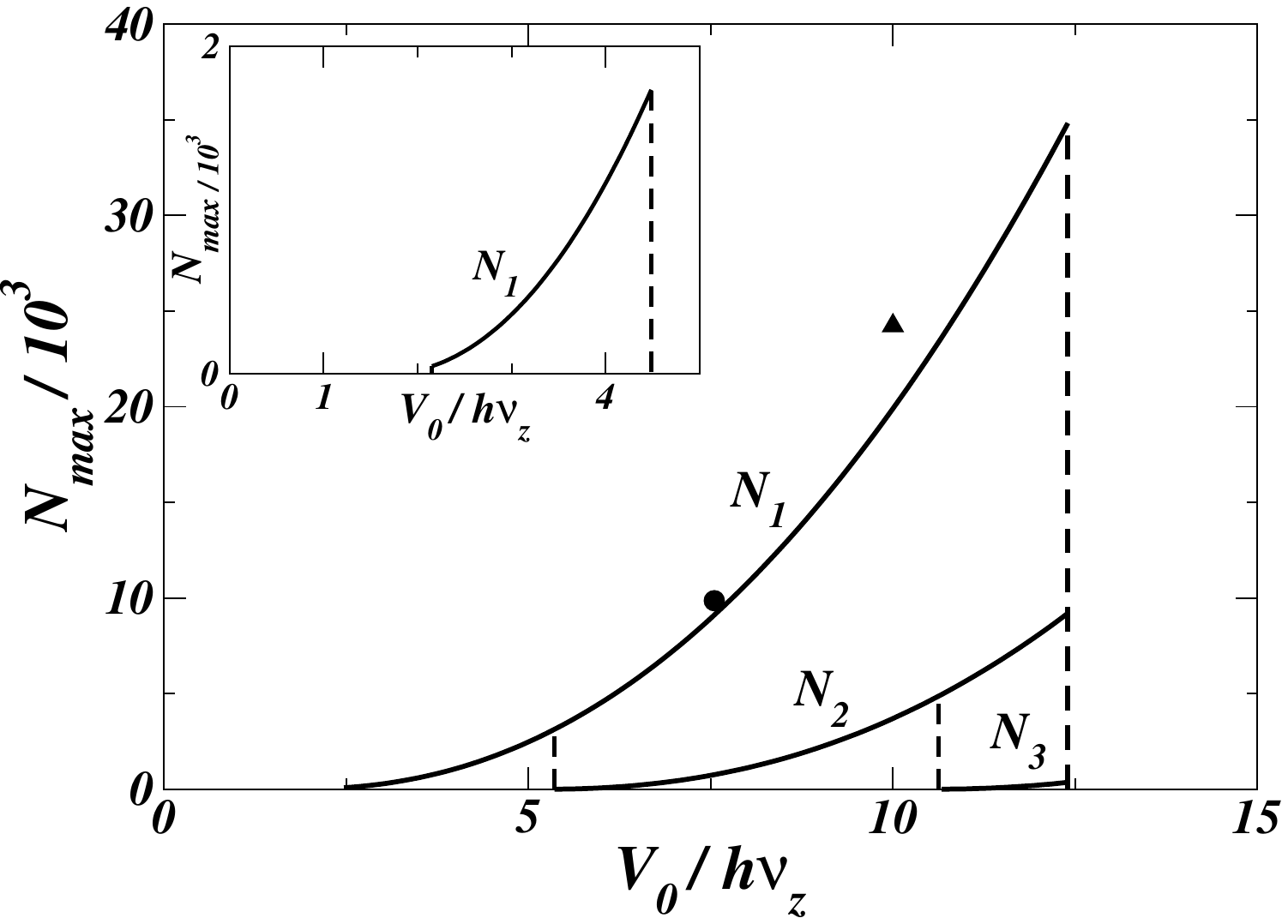} 
\caption{Plot of the maximum number of particles $N_1$, $N_2$ and 
$N_3$ which can be stored respectively in the first, 
second and third doublet as a function of the strength of the optical lattice
potential in Eq.(\ref{double_potential}) with 
$\nu_\perp= 10 Hz$, $\nu_z= 300 Hz$ and $A=0.15$. 
For these values of parameters up to three doublets 
have energy lower than $V_0$. The circle and the star correspond 
to $V_0/\hbar \omega_z=7.55$, $N=9.8 \cdot 10^3$ (circle) and 
$V_0/\hbar \omega_z=10$, $N=2.4 \cdot 10^4$ (triangle); in both cases 
the first and second doublet are completely occupied. 
Inset: maximum number of particles which can occupy 
the first doublet as a function of the strength of the optical lattice
potential Eq.(\ref{double_potential}) with $\nu_\perp= 10 Hz$, 
$\nu_z= 300 Hz$ and $A=0.25$.
For these values of parameters, there is only a doublet under barrier.}
\label{015} 
\end{figure}

We can estimate the number of particles which can be stored in each doublet
exploiting the degeneracy of the transverse degrees of freedom. Denoting 
with $\epsilon_{n}^{S,A}$ the symmetric 
and antisymmetric eigenvalues of the double well potential Eq.(\ref{double_potential}) 
corresponding to $n$-doublet and with $\Delta E_n$ 
the (average) energy difference between the $n+1$-th and the $n$-th doublet 
of the potential $V_{DW}(z)$:
\begin{equation}
\label{Delta_En}
\Delta E_n = 
\frac{(\epsilon_{n+1}^S+\epsilon_{n+1}^A)- (\epsilon_n^S+\epsilon_n^A)}{2}.
\end{equation}
Then the total number of particles which can be stored on one
side of the 3D double well (without occupying the next one) in the nth doublet is:
\begin{equation}\label{3D}
N_n^{3D} \approx \frac{1}{2} \left( \frac{\Delta E_n}{\hbar \omega_\perp}\right)^2.
\end{equation}
The number of particles in the lowest doublet is then given 
by $N_1^{3D} \approx \left( \Delta E_1/\hbar \omega_\perp\right)^2/2$: 
notice that for very large barriers, then 
$\Delta E_1 \sim \hbar \tilde{\omega}$ \cite{LANDAU81} 
(see also Appendix A).
If the energy of the higher doublet is 
beyond the energy of barrier ($s E_R$) then $\Delta E_n$ 
in Eq.(\ref{Delta_En}) must be replaced by 
$sE_R-\left( \epsilon_n^S+\epsilon_n^A \right) / 2$. 
From Fig.\ref{015} we observe that the number of particles in each doublet 
is much higher than in the 1D configuration, 
allowing to reach about $10^4$ particles when a few doublets are occupied. 
This allows the study tunneling phenomena with 
polarized fermions experimentally within this setting.

We conclude this Section by considering a 2D potential of the form
\begin{equation}
V_{2D}(x,z) = \frac{1}{2} m \omega_\parallel^2 x^2 + 
\frac{1}{2} m \omega_z^2 z^2 + V_0 \cos^2(k z):
\end{equation}
the number of particles which can be placed on a side of the double well 
in the $n$-th doublet is then given by
\begin{equation}
N_n^{2D} \approx \frac{\Delta E_n}{\hbar \omega_\parallel}
\end{equation}
with $\Delta E_n$ given by Eq.(\ref{Delta_En}). 
Of course, in the three-dimensional case 
the number of particles which can be stored is larger than the 2D double well 
potential due to the higher degeneracy introduced by the 
transverse degrees of freedom. 

\section{Tunneling dynamics in a double well potential: formalism}

In this Section we review the formalism to describe the dynamics of a system 
of polarized fermions in a 3D double well potential of the form 
Eq.(\ref{V_TOT}). The 1D limit, studied in \cite{SALASNICH10}, 
is retrieved and briefly discussed; we refer also to \cite{SALASNICH10} 
for a study of the effects of the boson-fermion interactions when
a localized Bose-Einstein condensates is placed in the wells.

The Hamiltonian for a system of polarized fermions confined 
in an external potential $V(\mathbf{x})$ reads
\begin{equation} \label{3DH}
H = \int d\mathbf{x}\  \psi^\dagger(\mathbf{x}) 
\left[-\frac{\hbar^2}{2m}\nabla^2 +V(\mathbf{x}) \right] \psi(\mathbf{x}).
\end{equation}
We denote with $\varphi_{n_x}(x)\varphi_{n_y}(y)\phi_\gamma(z)$ the eigenvectors of the single-particle Hamiltonian.
They are characterized by three quantum numbers: 
$n_x,n_y=0,1,\cdots$ denote the transverse quantum numbers associated 
to the transverse harmonic oscillator, 
while $\gamma$ denotes the eigenstates of the double well potential 
$V_{DW}(z)$, as obtained from
\begin{equation}
\left[-\frac{\hbar^2}{2m}\frac{\partial^2}{\partial z^2} +V_{DW}(z) \right]
\phi_\gamma(z)=
\epsilon_{\gamma} \phi_\gamma(z). 
\end{equation}
The quantum number $\gamma$ is defined by the pair $n,\alpha$, 
where $n=1,2,\cdots$ denotes the doublet, and $\alpha=S,A$ denotes 
the symmetric or antisymmetric state (in the following we denote 
$\epsilon_\gamma$ by $\epsilon_n^\alpha$). 
The eigenenergies of the single-particle Hamiltonian are given by 
$\varepsilon_{n_x n_y \gamma}= \hbar \omega_\perp(n_x+n_y+1) + 
\epsilon_\gamma$.

The fermionic field can be written as 
$\psi(\mathbf{x})=
\sum_{n_x n_y \gamma} \varphi_{n_x}(x)\varphi_{n_y}(y) \phi_\gamma(z) c_{n_x n_y \gamma}$. 
Since we want to study the dynamics of particles in the double well potential 
we decompose each pair of eigenfunctions 
corresponding to the same doublet into
its right and left component:
\begin{equation}
\psi(\mathbf{x})=
\sum_{n_x n_y n} \varphi_{n_x}(x)\varphi_{n_y}(y)
\left[\phi_n^R(z) c_{n_x n_y n}^R + \phi_n^L(z)  c_{n_x n_y n}^L \right]
\end{equation}
where the (Wannier) wavefunctions $\phi_n^{R,L}$ centered in the wells $R, L$ are given as usual by 
$$
\phi_n^{R}(z)=\frac{\phi_{n,S}(z) + \phi_{n,A}(z)}{\sqrt{2}}; 
\, \, \,
\phi_n^{L}(z)=\frac{\phi_{n,S}(z) - \phi_{n,A}(z)}{\sqrt{2}}.
$$
We will as well use the notations 
$\phi_{n_x n_y n,\alpha}(\mathbf{x})=
\varphi_{n_x}(x) \varphi_{n_y}(y) \phi_{n,\alpha}(z)$ 
for the eigenfunctions of the single-particle Hamiltonian 
and $\phi_{n_x n_y n}^{R,L}(\mathbf{x})=
\varphi_{n_x}(x) \varphi_{n_y}(y) \phi_n^{R,L}(z)$ for the 3D 
Wannier wavefunctions. 
The fermionic operators $c_{n_x n_y n}^{R,L}$ 
are defined as:
\begin{equation}
c_{n_x n_y n}^{R}= \frac{c_{n_x n_y n,S} + c_{n_x n_y n,A}}{\sqrt{2}};
\, \, \,
c_{n_x n_y n}^{L}= \frac{c_{n_x n_y n,S} - c_{n_x n_y n,A}}{\sqrt{2}}.
\end{equation}
The Hamiltonian (\ref{3DH}) then reads
\begin{eqnarray}
H &=&\sum_{n_x n_y n} \Bigg\{ 
\left( \hbar \omega_\perp \left(n_x+n_y+1 \right) + 
\frac{\epsilon_n^S+\epsilon_n^A}{2}  
\right) \cdot 
\left( c_{n_x n_y n}^{R\dagger} c_{n_x n_y n}^R + 
c_{n_x n_y n}^{L\dagger} c_{n_x n_y n}^L \right) + \\
&& + \frac{\epsilon_n^S- \epsilon_n^A}{2} \nonumber 
\left( c_{n_x n_y n}^{R\dagger} c_{n_x n_y n}^L + c_{n_x n_y n}^{L\dagger} c_{n_x n_y n}^R 
\right) \Bigg\}.
\end{eqnarray}
The density operator of particles along the direction of the 
double well is given by:
\begin{eqnarray}
n(z) = \int dx \ dy \ \psi^\dagger(\mathbf{x}) \psi(\mathbf{x})= \sum_{n_x n_y} \sum_{\gamma,\gamma'} 
\phi_{\gamma}^*(z)\phi_{\gamma'}(z) c_{n_x n_y \gamma}^\dagger c_{n_x n_y 
\gamma'}. 
\end{eqnarray}

Clearly, when we average the density operator over a generic state with a definite number of particles on the 
left and right part of the barrier the off-diagonal elements of the density operator give a null contribution.
We introduce the quantity 
$$\Delta N_{n_x n_y n}(t) = \left<c_{n_x n_y n}^{R\dagger} 
c_{n_x n_y n}^R \right> -  \left<c_{n_x n_y n}^{L\dagger} 
c_{n_x n_y n}^L \right>$$ which is the particle difference 
in the doublet $n_x,n_y,n$ between the right and the left sides of 
the double well for a generic initial state $\psi_0$ 
(given an operator ${\cal O}$, we use the notation $\left<{\cal O} \right>=\left<\psi_0 \mid {\cal O}(t) \mid \psi_0\right> $). Next we define the total 
population imbalance $\Delta N_n(t)$ 
for the n-th doublet:
\begin{equation}
\Delta N_n(t) = \sum_{n_x n_y} \left( \left<c_{n_x n_y n}^{R\dagger} 
c_{n_x n_y n}^R \right> -  \left<c_{n_x n_y n}^{L\dagger}
c_{n_x n_y n}^L \right>\right)
\end{equation} 
which evolves according to:
\begin{equation}
\frac{d^2}{dt^2} \Delta N_{n_x n_y n}(t) = -\left(\frac{\epsilon_n^A-\epsilon_n^S}{\hbar}\right)^2 \ \Delta N_{n_x n_y n}(t).
\end{equation}
This equation has the straightforward solution:
\begin{equation} \label{zn}
\Delta N_{n_x n_y n}(t) = \Delta N_{n_x n_y n}(0) \cos\left(2 \Omega_n t + \varphi_{n_x n_y n} \right), 
\end{equation}
where $\Omega_n= \frac{\epsilon_n^A-\epsilon_n^S}{2 \hbar} $ 
is the Rabi frequency for the oscillation of a particle into the same doublet 
and $\varphi_{n_x n_y n}$ depend in general on the initial conditions. 

The total \textit{fractional}  population imbalance $z(t)$ is defined as
\begin{equation}
z(t) = \frac{\sum_{n} \Delta N_n(t)}{N}.
\end{equation}
For a single particle in the doublet $n_x, n_y, n$, 
the initial state $\psi_0(\mathbf{x})$ can be in general be written 
in the Bloch sphere as 
\begin{equation}
\psi_0(\mathbf{x}) = \cos\left(\frac{\theta_{n_x n_y n}}{2}\right) 
\phi_{n_x n_y n,S}(\mathbf{x})+  
 e^{i\varphi_{n_x n_y n}}\sin\left(\frac{\theta_{n_x n_y n}}{2}\right) \phi_{n_x n_y n,A}(\mathbf{x}),
\end{equation}
where $\theta_{n_x n_y n}$ and $\varphi_{n_x n_y n}$ are the coordinates on the Bloch sphere: 
when the particle is initially in the state $\phi_{n_x n_y n}^R(\mathbf{x})$ ($\phi_{n_x n_y n}^L(\mathbf{x})$), then $\theta_{n_x n_y n}=\pi/2$ and $\varphi_{n_x n_y n}=0$ 
($\varphi_{n_x n_y n}=\pi$). The energy of such a state is equal to:
\begin{equation}
\left< \psi_{0} |H|\psi_{0}\right> = 
\epsilon_n^S \cos^2\left(\frac{\theta_{n_x n_y n}}{2}\right) + \epsilon_n^A 
\sin^2\left(\frac{\theta_{n_x n_y n}}{2}\right) .
\end{equation}
It is then easy to see that the population imbalance for this state evolves according to (\ref{zn}) where
\begin{equation}
\Delta N_{n_x n_y n}(0)= \sin{\theta_{n_x n_y n}}.
\end{equation}
 
Let consider now many fermions in the double well potential. 
For a generic initial state with the fermions in the same n-th doublet 
with different initial phases $\varphi_{n_x n_y n}$, 
then oscillations in the population imbalance are rapidly washed out: indeed, 
since the fermions are polarized, the dynamics in different 
doublets are independent and dephasing of the different 
particle differences $\Delta N_{n_x n_y n}(t)$ in general occurs. 
However the experimental situation we have in mind is a system of two 
identical wells initially practically decoupled (i.e., with a very high 
energy barrier between them) with a
definite number of particles (say $N_L$ and $N_R$, with $N_L<N_R$) in each side in their ground state. 
Then, at the initial time $t=0$ the barrier along the $z$ direction is lowered 
and the system is left to evolve freely with the Hamiltonian (\ref{3DH}). 
Alternatively, one could start from the ground state of a symmetric 
3D double well and tilt for some time the double well, 
so that the energy minima 
of the double well potential are different and particles flow toward 
the energetically favoured well: removing abruptly the tilted 
potential one has an initial state with a different number of particles 
in the two wells. 

For sufficiently high barriers the first $N_L$ particles will be frozen 
in the two wells, 
while the others will start to tunnel. 
Take as initial state for one of such particles 
in the right side $\psi_{n_x n_y n_z}$ where 
$n_x$ and $n_y$ are the quantum numbers along 
the transverse directions which remain unperturbed and 
$n_z$ labels the states along the longitudinal $z$-axis 
before the barrier gets lowered;
then the initial state  $\psi_0(\mathbf{x})=\psi_{n_x n_y n_z}(\mathbf{x})$ can be well approximated by
\begin{equation}
\psi_{0}(\mathbf{x}) \approx \phi_{n_x n_y n}^{R}(\mathbf{x}).
\end{equation}
Therefore we decide to study the evolution of the system of an arbitrary number of fermions under the conditions that the 
initial state of each particle is of the type 
$\varphi_{n_x}(x)\varphi_{n_y}(y)\frac{\phi_n^S(z)\pm \phi_n^A(z) }{\sqrt{2}}$ to have a reasonable description of the dynamics.
 
If, for instance, we suppose all the particles in the n-th doublet 
initially on the
right well the phases $\varphi_{n_x n_y n}$ can be taken equal to zero and the number of particles in each side of the 
barrier read:
\begin{equation}
n_n^R \equiv \sum_{n_x n_y} \left<c_{n_x n_y n}^{R\dagger} c_{n_x n_y n}^R 
\right> 
= \Delta N_n(0) \cos^2(\Omega_n t)
\end{equation}
\begin{equation}
n_n^L \equiv \sum_{n_x n_y} \left<c_{n_x n_y n}^{L\dagger} c_{n_x n_y n}^L 
\right> 
= \Delta N_n(0) \sin^2(\Omega_n t).
\end{equation}

We observe that in the 1D limit the transverse degrees 
of freedom are frozen and fermions belonging 
to different doublets oscillates with different frequencies: this case 
has been 
studied in detail in \cite{SALASNICH10}. Oscillations follow an 
aperiodic time dependence due to the superposition of different incommensurate 
oscillations: for typical experimental values one sees that already for 
hundreds of particles the current oscillations 
are practically washed out for an initial imbalance not too small.
Moreover, since Rabi frequencies for distant levels can be very different, 
the time scale required for a complete oscillation of the population imbalance $z(t)$ can be much larger than the experimental observation times. 

\section{Dynamics in a three-dimensional double well potential}

In this Section we consider the dynamics of polarized fermions 
in the 3D confining potential (\ref{V_TOT}) when the 1D validity condition 
is violated, as it happens for realistic transverse confining potentials. 
As discussed in Section II, the number of particles which can be stored in one
doublet can be very much increased by taking a weak confinement in
the transverse direction 
thanks to the possibility to fill the transverse states before 
reaching the next doublet. In the inset of Fig.\ref{015} 
we consider the range of parameters $V_0^{(min)}<V_0<V_0^{(max)}$ 
such that the double well potential (\ref{V_DW}) 
has only a doublet under barrier and we plot in this range  
the maximum number $N_\text{max}$ of particles which can be stored in that 
(first) doublet: in other words, for a number of atoms $N>N_\text{max}$ they 
start to occupy levels above the top of the barrier. One sees 
that $N_\text{max} \sim 10^3$ 
for reasonable experimental parameters of the potential. 
Since in this case there is only one 
doublet with energy below $V_0$ then the number of particles
scales as $V_0^2$ (the energy of the particle in the double well direction, 
$\frac{\epsilon_1^S+\epsilon_1^A}{2}$, grows as $\sqrt{V_0}$).  

The number of particles can be further increased 
by decreasing the ratio $A=\frac{a_{ho}}{\lambda}$, 
as shown in Fig.\ref{015}. In this figure we consider 
values of the parameters such that the double well potential 
(\ref{V_DW}) can have one, two or three doublets: we plot the maximum 
number of particles $N_n$ which can be stored in the doublet $n=1$, $2$ or $3$. 
The maximum number of particles in the first 
level can arrive to $N_\text{max} \approx 3.5\cdot 10^4$.

\begin{figure}
\centering
\includegraphics[width=7.cm,angle=0,clip]{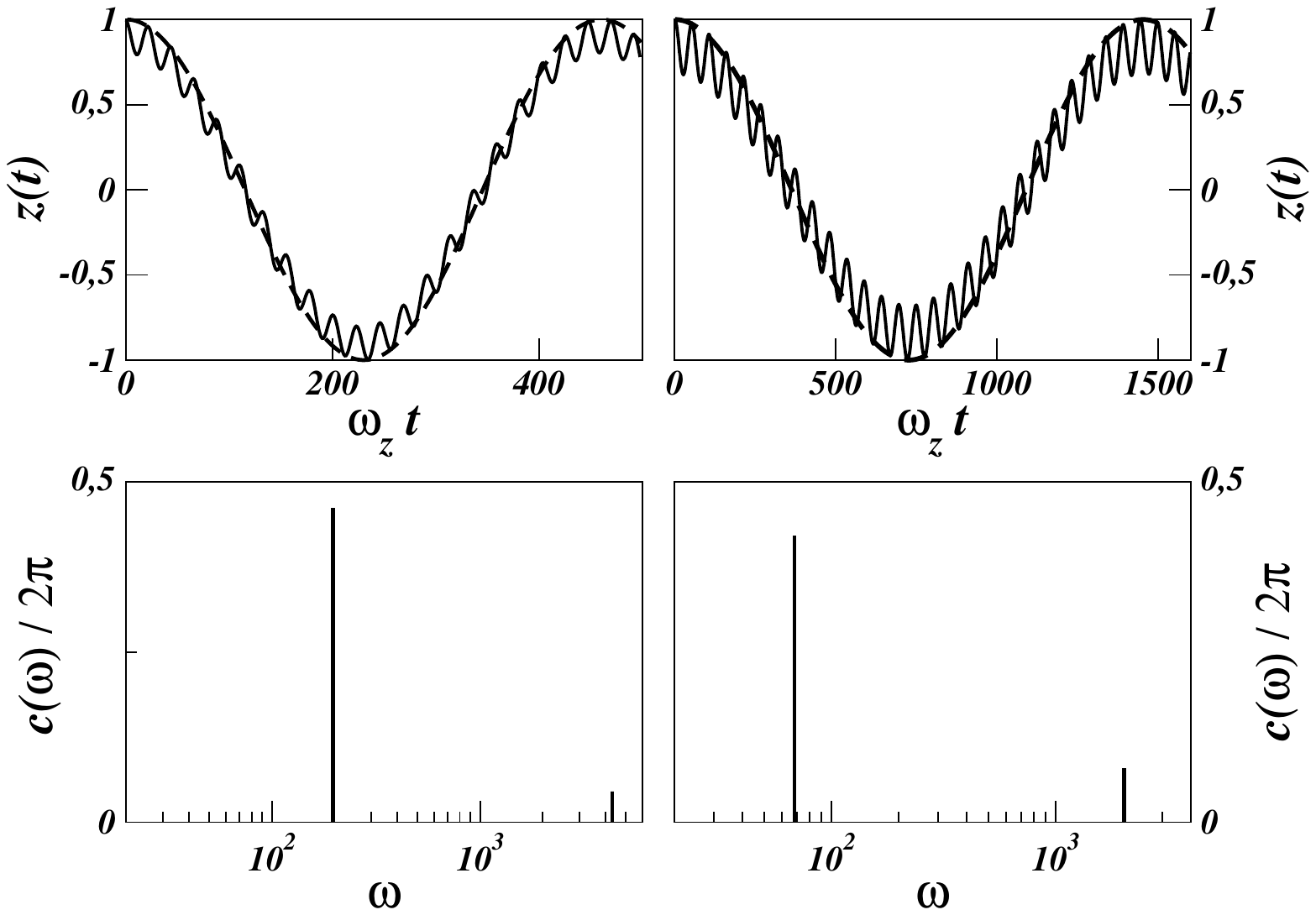} 
\caption{Above: (full line) evolution of $z(t)$ for the points of Fig.\ref{015}: the circle (left) and the up triangle
(right); (dashed line) evolution for a single particle in the lowest doublet with equal initial conditions. Below: 
magnitude of the modulus $c(\omega)$ of the Fourier
transform of $z(t)$  for the two plots above.}
\label{three_dynamics}
\end{figure}

\begin{figure}[ht]
\begin{minipage}[t]{0.4\linewidth}
\centering
\includegraphics[width=6cm,angle=0,clip]{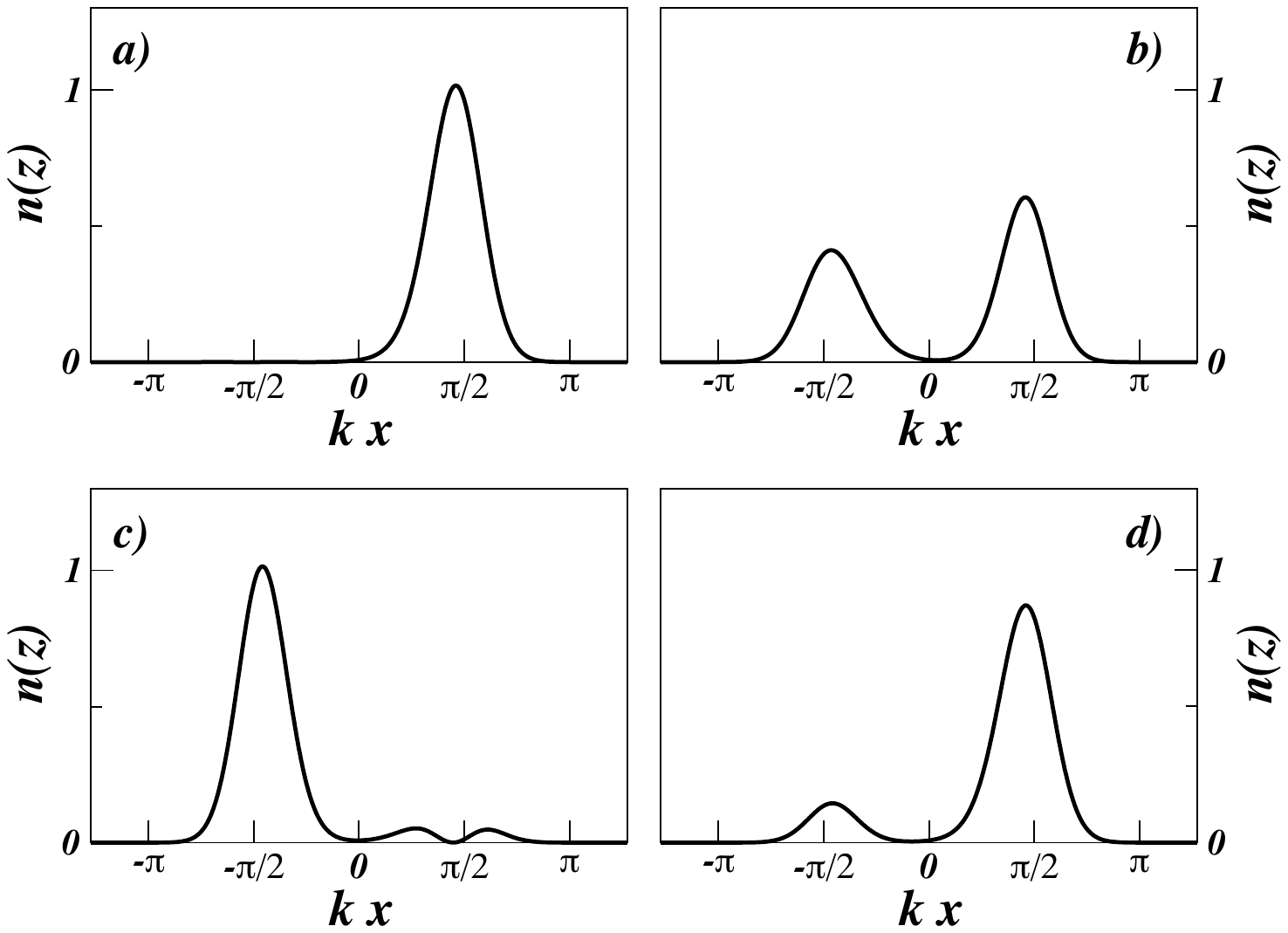} 
\caption{From above left: normalized particle density at different times for the circle of Fig.\ref{015}. 
We take $9.8\cdot 10^3$ particles distributed among the
first two doublets. a), b), c), d): $\omega_z t = 0, 100, 225, 400$. }
\label{dynamics_circle}
\end{minipage}
\hspace{0.5cm}
\begin{minipage}[t]{0.4\linewidth}
\centering
\includegraphics[width=6cm,angle=0,clip]{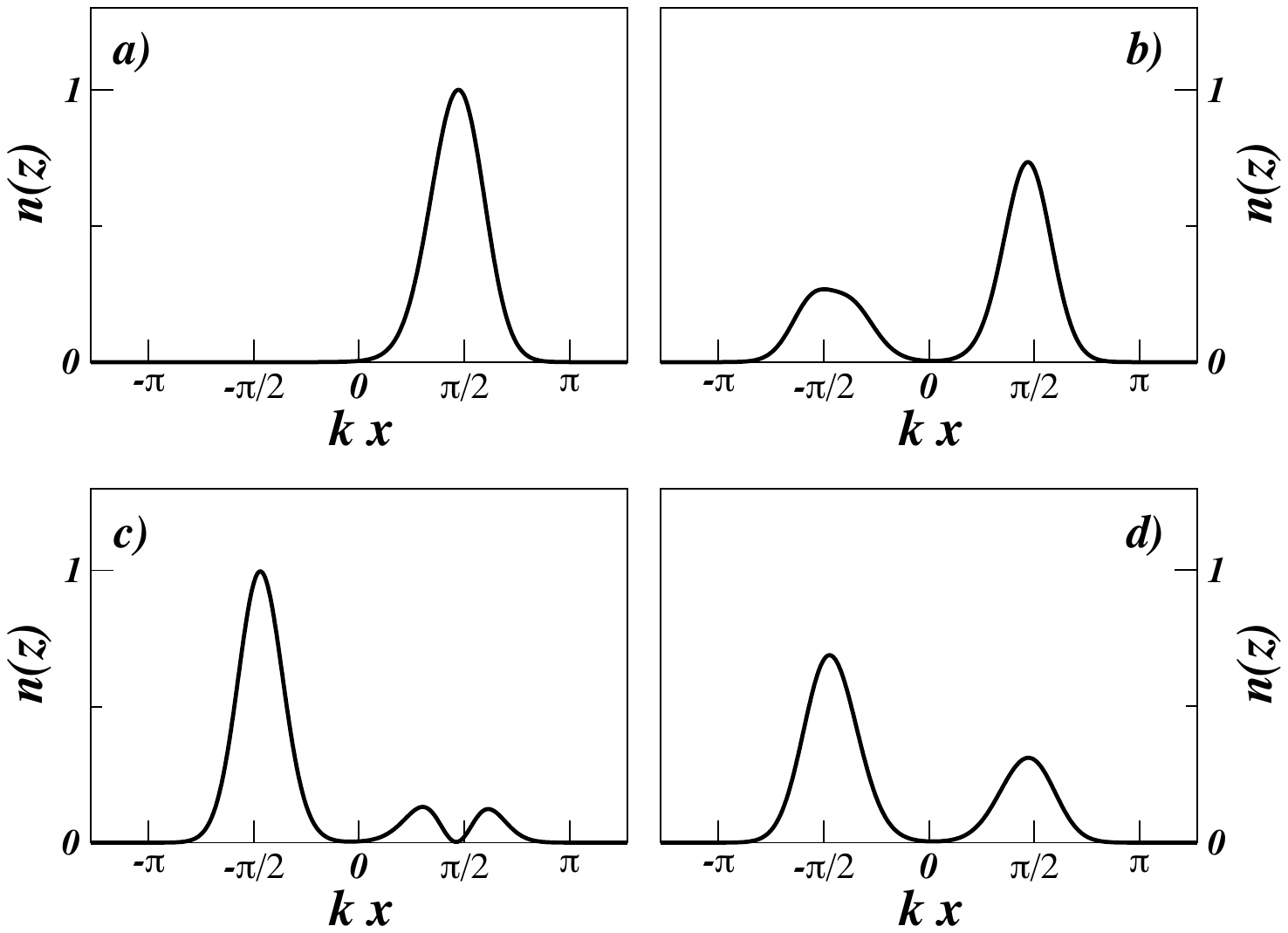} 
\caption{From above left: normalized particle density at different times for the triangle of Fig.\ref{015}. 
We take $N=2.4\cdot 10^4$ particles distributed among the
first two doublets. a), b), c), d): $\omega_z t = 0, 250, 750, 1000$. }
\label{dynamics_triangle}
\end{minipage}
\end{figure}

The dynamics for the fractional population imbalance $z(t)$ 
is illustrated in Fig.\ref{three_dynamics};
here for simplicity we show the results for the evolution of particles initially distributed in the
the right side of the barrier.
In the left 
part of the figure we considered $N=9.8\cdot 10^3$ atoms 
distributed among the first two doublets filling completely
the first one and almost totally also the second at 
$V_0=7.5 \hbar \omega_z$ (corresponding to the circle in Fig.\ref{015}): 
we see that the oscillation of $z(t)$
contains only two Rabi frequencies. The (normalized) 
particle density at different times is plotted in Fig.\ref{dynamics_circle}. 
In the right part of Fig.\ref{three_dynamics} we considered 
$N=2.4\cdot 10^4$ atoms distributed among the first 
(completely filled) and the second doublet at $V_0=10 \hbar\omega_z$ 
(corresponding to the triangle in Fig.\ref{015}), and the 
particle density at different times is plotted in Fig.\ref{dynamics_triangle}.  In this case, 
the profile of the densities has more pronounced secondary peaks 
due to the contribution coming from particles in 
the second doublet.

Finally we observe that a way to observe 
the relative number of particles in each doublet at the initial
time $t=0$ is through the analysis of the Fourier transform 
of the fractional population imbalance.
Since our considerations are done at $T=0$ 
the motion of the particles in the system is undamped, so
the Fourier transform of $z(t)$ is simply given by a sum of 
$\delta$-functions, each one localized at the
characteristic Rabi-frequency of each doublet: for the cases 
considered in Figs.\ref{three_dynamics}-\ref{dynamics_triangle}, where 
only two doublets are occupied, one has
\begin{equation}
z(\omega)= \int_{-\infty}^{+\infty} e^{i \omega t} z(t) dt = 2\pi \left(\frac{N_1}{2N}\left[ \delta(\omega-\Omega_1)+ \delta(\omega+\Omega_1)\right]+
		            \frac{N_2}{2N}
\left[ \delta(\omega-\Omega_2)+ \delta(\omega+\Omega_2)\right]\right)	
\end{equation}
where $N_1$ and $N_2$ are the number of particles respectively 
in the first and second doublet. In the bottom of
Fig.\ref{three_dynamics} we plot the coefficients of the Fourier transform for the two cases described above.

To conclude this Section, we mention that it would be interesting 
to study in the future 
the strongly driven dynamics of polarized fermions in 3D double wells 
in presence of time-dependent potentials \cite{ECKARDT05,BURENKOV10}, 
as well as their dynamics 
in presence of interparticle (eventually repulsive) interactions 
in order to investigate in this experimental setup the issues 
of equilibration and thermalization 
\cite{CALABRESE07,IUCCI09,SOTIRIADIS09,FIORETTO10,GENWAY10,BIROLI10,YUKALOV10,YUKALOV11,BANULS11}.

\section{Conclusion}
We studied the tunneling dynamics of a spin polarized Fermi gas in 
a three-dimensional double well potential at zero temperature. We focused in particular 
on the time dynamics starting from an initial state 
in which there is an imbalance 
between the number of particles in the two wells. 
Although fermions in different 
doublets of the double well tunnel with different frequencies, we point out 
that (incoherent) oscillations of a large number of particles can arise, 
as a consequence of the presence of transverse degrees of freedom. 

Estimates of the 
doublet structure and of the occupation of transverse eigenstates 
for a realistic experimental setup are provided. In the 
1D limit the current oscillations are washed out, as a result of the 
dephasing, but for not too large confining transverse frequencies or not 
too small number of particles the fermions can occupy only the 
first doublets (using the transverse eigenstates), resulting in oscillations 
of a large number of fermions. 
We stress that these are incoherent single-particle oscillations. 
We can conclude that if the 1D condition is violated, then the simple 
observation of a sinusoidal current cannot in general simply discriminate between 
incoherent and coherent (Josephson) tunneling dynamics at zero temperature.

{\it Acknowledgements:} Discussions with L. Salasnich, A. Smerzi, 
S. Chiacchiera, A. Recati, G. Roati, G. Gori and M. Iazzi  
are warmly acknowledged; we also thank L. Pezz\`e for useful 
correspondence. 
This work is supported by the grants INSTANS (from ESF) and 
2007JHLPEZ (from MIUR).

\appendix

\section{Semiclassical estimation of the splittings}

In this Appendix we report a semiclassical estimation of the 
energy splitting among the levels of the
double well potential (\ref{double_potential}). 
The semiclassical formula for the splitting of the n-th energy level is 
given by \cite{LANDAU81}
\begin{equation}
E_n^A-E_n^S =
\frac{\hbar \omega_{cl}}{\pi} e^{-\frac{1}{\hbar} \int_{-a}^a dx\ |p| },
\end{equation}
where $a, -a$ denote the turning points 
at the central barrier of the potential 
(classical motion is inhibited there), $|p| = \sqrt{2m(V(z)-E)}$
and $\omega_{cl}$ is the frequency of the classical motion between 
the turning points where
classical motion is allowed.

In the Tab.\ref{table1} we compute the semiclassical splitting energy in two cases: in the computation of $E^I_\text{semi}$ 
we used the estimation for the average energy $E= \hbar \tilde\omega \left( n+ \frac{1}{2} \right)$ of the doublet
based on the effective frequency (\ref{omega_tilde}), in $E^{II}_\text{semi}$ 
we used instead the average value of the energy 
of each doublet obtained from the numerical solution of the Schr\"odinger equation for the potential (\ref{double_potential}).
As we might expect the agreement of the semiclassical estimation with the numerical values is much better in the second case; we
also note that in the first one there is a good agreement (the error is around $10\%$) for the first two doublets due to the validity
of using the effective frequency (\ref{omega_tilde}) for the lowest energy levels.

\begin{table}
\begin{tabular}{c|c|c|c|c|c}
Doublet & $\Delta E_{\text{Exact}}$ & $\Delta E^I_{\text{semi}}$ & $\frac{E^I_\text{semi}-E_\text{Exact}}{E_\text{Exact}}$ & $\Delta E^{II}_{\text{semi}}$ & $\frac{E^{II}_\text{semi}-E_\text{Exact}}{E_\text{Exact}}$\\ 
\hline
1 		 &  	$5.44\cdot10^{-5}$	& 	$4.95\cdot10^{-5} $ 	    &	$-0.09$	& 	$4.93\cdot10^{-5} $ 	    &			$-0.094$	\\
2 		 & 	$3.34\cdot10^{-3}$	& 	$3.70\cdot10^{-3} $	    &	$ 0.11 $	& 	$3.19\cdot10^{-3} $	    &			$ - 0.045$	\\
3 		 & 	$8.75\cdot10^{-2}$	&     $0.142$			    &	$0.62 $	&     $8.85\cdot10^{-2}$	    &			$ 0.011$\\
4 		 &	$1.12$				& 	$2.80$			    &	$1.50$	& 	$1.14$			    &			$0.018$\\
\hline
\end{tabular}
\caption{Semiclassical estimation of the splitting energy using the value given by (\ref{omega_tilde}) for the energy $E$ ($E^I$) or the numerical value for the average energy of the doublet ($E^{II}$). 
Energies are in units of $E_R$. We use here $s=76$ and $A=0.1$ corresponding to four doublets.}
\label{table1}
\end{table}


\begin{thebibliography}{*}
\bibitem{NOTE1}
The first two minima of the potential 
(\ref{V_DW}) which identify the position of the
double well minima are given by the solution of the transcendental equation
$\sin{(2kz)}/2kz = 1/(2 \pi A)^4  s$.
\end{thebibliography}
\end{document}